# Unexpected two-fold symmetric superconductivity in few-layer NbSe₂

Alex Hamill[1†], Brett Heischmidt[1†], Egon Sohn[2,3†], Daniel Shaffer[1], Kan-Ting Tsai[1], Xi Zhang[1], Xiaoxiang Xi[4], Alexey Suslov[5], Helmuth Berger[6], László Forró[6], Fiona J. Burnell[1], Jie Shan[3,7], Kin Fai Mak[3,7], Rafael M. Fernandes[1], Ke Wang[1*] and Vlad S. Pribiag[1*]

[1]School of Physics and Astronomy, University of Minnesota, Minneapolis, MN, USA

[2]Department of Physics, The Pennsylvania State University, University Park, PA, USA

[3]Department of Physics and School of Applied and Engineering Physics, Cornell University, Ithaca, NY, USA

[4]National Laboratory of Solid State Microstructures, School of Physics, and Collaborative Innovation Center of Advanced Microstructures, Nanjing University, Nanjing, China

[5]National High Magnetic Field Laboratory, Tallahassee, FL, USA

[6]Institute of Condensed Matter Physics, Ecole Polytechnique Fédérale de Lausanne, Lausanne, Switzerland

[7]Kavli Institute at Cornell for Nanoscale Science, Ithaca, NY, USA

**Two-dimensional transition metal dichalcogenides (TMDs) have been attracting significant interest[1–8] due to a range of properties, such as layer-dependent inversion symmetry, valley-contrasted Berry curvatures, and strong spin-orbit coupling (SOC). Of particular interest is niobium diselenide (NbSe₂), whose superconducting state in few-layer samples is profoundly affected by an unusual type of SOC called Ising SOC[7]. Combined with the reduced dimensionality, the latter stabilizes the superconducting state against magnetic fields up to ~35 T and could lead to other exotic properties such as nodal and crystalline topological superconductivity[9–14]. Here, we report transport measurements of few-layer NbSe₂ under in-plane external magnetic fields, revealing an unexpected two-fold rotational symmetry of the superconducting state. In contrast to the three-fold symmetry of the lattice, we observe that**

*Corresponding authors: vpribiag@umn.edu, kewang@umn.edu  
† These authors contributed equally

**the magnetoresistance and critical field exhibit a two-fold oscillation with respect to an applied in-plane magnetic field. We find similar two-fold oscillations deep inside the superconducting state in differential conductance measurements on NbSe$_2$/CrBr$_3$ superconductor-magnet junctions. In both cases, the anisotropy vanishes in the normal state, demonstrating that it is an intrinsic property of the superconducting phase. We attribute the behavior to the mixing between two closely competing pairing instabilities, namely, the conventional s-wave instability typical of bulk NbSe$_2$ and an unconventional d- or p-wave channel that emerges in few-layer NbSe$_2$. Our results thus demonstrate the unconventional character of the pairing interaction in a few-layer TMD, opening a new avenue to search for exotic superconductivity in this family of 2D materials.**

In the conventional BCS theory of superconductivity, the pairing state is characterized by the breaking of an U(1) symmetry related to fixing the phases of the Cooper-pairs wave-function. Yet, several superconductors break additional symmetries, leading to new forms of superconductivity. Symmetry breaking in a superconductor may arise through intrinsically anisotropic pairing channels, as well as through coupling between different channels promoted by external perturbations. Hexagonal NbSe$_2$ is a member of the transition metal dichalcogenide (TMD) family that exhibits superconductivity[15] and charge-density wave order[16–18] from bulk to monolayer forms. In bulk, NbSe$_2$ is believed to be in the s-wave pairing state[19]. However, in few-layer hexagonally-stacked 2H-NbSe$_2$, the underlying crystalline symmetries give rise to distinct electronic properties. For example, odd-layer-number NbSe$_2$ lacks inversion symmetry (Figures 1a,b), which leads to a spin-orbit interaction that polarizes the spins of different valleys in different out-of-plane directions[7], as illustrated by the Fermi surface plotted in Figure 1c. This so-called Ising-SOC stabilizes the superconducting state against in-plane magnetic fields far exceeding the Pauli paramagnetic limit imposed by conventional superconducting theory[20,21]. This effect becomes stronger as the monolayer limit is approached[7,22]. In addition to Ising superconductivity[9–11], other unique superconducting properties such as field-induced mixed-parity states[23,24] and nodal topological superconductivity have been theoretically proposed[12–14,25]. However, other than the stability of the superconducting state against high fields, little is known experimentally about the nature and symmetry of the pairing state in few-layer $NbSe_2$.

In this work, we report two-fold-periodic superconducting properties of few-layer encapsulated $NbSe_2$ samples studied under rotating in-plane magnetic fields. This periodicity is confirmed by three complementary experimental approaches: magneto-transport near the transition temperature ($T_c$), measurements of the critical field ($H_c$), and tunneling across magnetic tunnel junctions deep inside the superconducting state. The superconducting properties follow a $\cos(2\theta)$ dependence on the in-plane field angle as probed with all three experimental approaches. These results are suggestive of field- or strain-induced mixing between quasi-degenerate pairing states with s-wave and d- or p-wave symmetries in few-layer samples, thus shedding new light on the unusual superconducting state of few-layer TMDs.

Figure 1d shows Device 1 (schematic in Figure 1e), an hBN-encapsulated five-layer $NbSe_2$ sample, used for magneto-transport measurements. Figure 1g shows Device 2 (schematic in Figure 1h), a magnetic tunnel junction, which consists of a normal metal (Pt), bilayer $CrBr_3$ and trilayer $NbSe_2$. $CrBr_3$ with H-stacking is a magnetic semiconductor[26,27] (gap 0.6-3.8 eV[28]) with a ferromagnetic ordering temperature of 36 K[29] and a magnetic anisotropy field (as reflected by the in-plane saturation field) ~440 mT[29–31]. For both types of devices, the $NbSe_2$ was exfoliated and encapsulated (see Methods for details) in Ar-filled gloveboxes (<0.1 ppm $O_2$ and $H_2O$ concentration) to preserve a pristine nature for the $NbSe_2$ under study. The same approach was also followed for $CrBr_3$ in the case of the magnetic tunnel junctions.

Figure 1f shows the temperature-dependence of the resistance of Device 1. The sample shows a clear drop from the normal state resistance ($R_N$) to vanishing resistance, with a superconducting transition temperature $T_c \approx 5.0$ K (defined as the point where $R = 0.95 R_N$). This value of $T_c$ is consistent with the value of the superconducting gap extracted from tunneling data on 2D NbSe$_2$[32]. The device shows similar transport characteristics as those studied in Ref. 7: phonon-limited linear resistance ($R \propto T$) for high temperatures, indicating a metallic sample, and disorder-limited transport (R approaching a plateau) just above the superconducting transition. $R_N$ is defined as the resistance value at this plateau. The resistance remains low (23.6 Ω) just above the superconducting drop, indicating the sample remains metallic all the way to the superconducting transition. Figure 1i shows similar four-probe data for the magnetic junction sample (Device 2). The four-probe resistance of the NbSe$_2$ flake measured as a function of temperature on a section non-overlapping with the CrBr$_3$ flake shows a superconducting transition with $T_c \approx 5.5$ K. The zero-bias differential conductance of the junction shows a similar onset temperature; however, the junction resistance saturates towards a finite value of 0.6 $R_N$ due to the finite resistance of the CrBr$_3$ bilayer in series with the NbSe$_2$. The inset of Figure 1i shows the differential conductance (G) of the junction vs. d.c. bias at T=1.9 K. We fit the spectrum with the Blonder-Tinkham-Klapwijk (BTK) model, which considers both quasiparticle tunneling and Andreev reflection processes[33,34]. The experimental data agree well with the fit except for the dip observed outside the superconducting gap. The small dip can be attributed to local heating in junctions with high transparency[35]. The superconducting gap extracted from this fit, of about 0.67 meV, is comparable to the $T_c$ value extracted from the resistance measurements.

Next, we discuss the magneto-transport properties of Device 1, probed under an in-plane magnetic field. Care was taken to rule out effects of an accidental out-of-plane component of the field (see Supplementary Sections 1, 2, 3, and 5). As the magnetic field is rotated in-plane (θ = 0 corresponds to magnetic field aligned along the NbSe$_2$ flake's longest straight edge), we observe a pronounced two-fold modulation of the magnetoresistance in the range between the onset and the offset of superconductivity (Figures 2a-e), which is consistent across multiple samples. Such a modulation is well described by a sinusoidal function of the form $\cos(2\theta + \varphi)$, as shown by the solid lines. As we varied the temperature for a fixed applied field amplitude of 8T, the angular modulation of the resistance ($\Delta R$) was suppressed when the resistance was outside of the superconducting transition region (see Figure 2d). The phase of the oscillating signal also shifted slightly in Device 1 as a function of field and temperature (most apparent in Figures 2b and 2e). The field angles at which the resistance maxima or minima were observed were not affected by the directions of voltage measurement, current, and material transfer. However, the field angle of minimum resistance does show a consistent alignment with the long straight edge of the crystal (see Supplementary Section 5). We expect the long straight edge may be associated with the direction of cleaving during exfoliation, which may align along the zigzag ($\vec{a}$) or armchair ($\vec{b}$) directions.

The measurements described above were performed around $T_c$, in a temperature regime where superconducting fluctuations are present. To verify that this effect persists at lower temperatures, a similar 5-layer device (Device 3, $T_c \approx 6.4K$) was studied at a temperature of 0.5 K in fields up to ~35T, revealing a similar modulation of the magneto-resistance with applied field angle (Figure 2c and Supplementary Figure S4). For this device, we also studied the angular dependence of the critical field $H_c$ that suppresses superconductivity, defined as the field at which the resistance is $0.5\ R_N$ (measured at $T = 0.5K$). $H_c(\theta)$ also showed two-fold periodicity (Figure 2c). The oscillations of $H_c(\theta)$ and the magnetoresistance have a $\pi$ phase shift, such that at angles where superconductivity is hardest to suppress, $H_c$ is largest and the magnetoresistance is lowest, as expected. Importantly, these data indicate that the angle-dependence of the magnetoresistance is a good proxy for the angle-dependence of $H_c$.

To examine the possibility that small out-of-plane field contributions arising from misalignment could lead to these observations, the magneto-transport measurements were repeated in a canted configuration where the sample was intentionally canted by ~7-9 degrees out of the field rotation plane. The magnetoresistance oscillations in the canted configuration show a $|\sin(\theta)|$ form, as expected from geometric considerations, which is in contrast with the $\cos(2\theta + \varphi)$ dependence reported in the absence of canting (see Supplementary Section 1).

To confirm that the two-fold anisotropic features also emerge deep inside the superconducting state, we also studied the NbSe$_2$/CrBr$_3$ junction (Device 2). In particular, we measured the differential conductance spectra under a 3T in-plane magnetic field as a function of the angle θ, defined in the same manner as for Device 1. We also studied a second type of junction with a ferromagnetic (Co/Pt multilayer) electrode (Device 4) (see Supplementary Section 7). At 3T, the value of the applied field is expected to be substantially larger than the in-plane saturation field (~440 mT)[29] for few-layer CrBr$_3$, ensuring that its spins are oriented in the plane. Figure 3a summarizes the spectra in a color plot and shows a clear two-fold modulation, especially prominent at low bias smaller than or comparable to the superconducting gap. The angular dependence of the differential conductance at a fixed bias shows a clear contrast between biases larger or smaller than the gap (Figure 3b). The normalized differential conductance at energies outside the gap (V=4mV > Δ/e) has less than a 0.5% variation vs. angle, indicating negligible anisotropic differential conductance. In contrast, the angular dependence of the normalized conductance at energies inside the gap (V = 0 mV) shows a clear two-fold modulation which can be fit to a $\cos(2\theta + \varphi)$ form (solid curve in Figure 3b). Overall, this indicates that the two-fold modulation persists deep inside the superconducting state of NbSe$_2$, consistent with the magneto-resistance and H$_c$ behavior.

The differential conductance spectra at θ = 45° and 135° (Figures 3c,d) show the variation of the angular dependence vs. field amplitude. At H=0, we observe a large differential conductance at low bias, approaching 1.6 times the normal state value. This is due to a combination of Andreev reflection and quasiparticle tunneling at the junction. When a finite magnetic field is directed along θ = 45°, the differential conductance becomes greatly suppressed. In contrast, when the magnetic field is directed along θ = 135°, the conductance does not change appreciably compared to that of 0 T, indicating a higher resilience of the superconducting state to magnetic fields at this angle, and highlighting the two-fold anisotropy of this state. Note that due to the ferromagnetism of $CrBr_3$, we expect magnetic proximity to be present at the $CrBr_3$/$NbSe_2$ interface[36], enhancing the effect of the applied magnetic field. In previous studies by some of us[34], we have investigated over ten bilayer and trilayer $NbSe_2$ tunnel junctions with nonmagnetic insulating barriers and electrodes under an arbitrary in-plane magnetic field direction relative to the $NbSe_2$ edge and have observed a monotonically decreasing gap from zero field up to the upper critical field (~38T). In all devices, the spectra evolved in a manner qualitatively identical to that in Figure 3d, implying there is no strong angle dependence of the tunneling spectra without magnetic barriers and electrodes. Similarly, any out-of-plane magnetic field due to misalignment would also lead to a monotonically decreasing gap and therefore cannot result in the large suppression of zero bias differential conductance and widening of the spectra under magnetic field observed in this study (Figure 3c). The strong two-fold angular periodicity of the tunneling spectra is unique to devices with a magnetic insulating barrier (Device 2) or a magnetic electrode (see Supplementary Section 7).

Having experimentally established the two-fold anisotropic character of the superconducting state of few-layer $NbSe_2$, we now discuss its possible origin. One scenario is that the superconducting state spontaneously breaks the three-fold rotational symmetry of the lattice, i.e. it is a nematic pairing state, similar to that observed in doped $Bi_2Se_3$[37]. This is only possible if two conditions are satisfied: (i) the gap function transforms as the E' or E'' irreducible representations of the $D_{3h}$ point group associated with the single-layer $NbSe_2$ crystal structure (which correspond roughly to d-wave and p-wave gaps, respectively), and (ii) the d-wave/p-wave nematic gap configuration has a lower energy than the d-wave/p-wave chiral gap configuration. However, the facts that the superconducting state of bulk NbSe$_2$ is s-wave[19] (i.e. it transforms as the A$_1$' irreducible presentation of D$_{3h}$), and that T$_c$ seems to continuously and slowly change as a function of the number of layers[7], pose a significant challenge to this scenario.

How could then an s-wave pairing state display two-fold anisotropy? One interesting possibility is that a small external symmetry-breaking field induces a strong mixing between the leading s-wave (A$_1$') instability and a sub-leading instability with either d-wave (E') or p-wave (E'') character. Such a strong mixing is of course only possible if these sub-leading instabilities of unconventional character, which presumably are accentuated in few-layer NbSe$_2$ due to the prominent role played by Ising SOC, are close competitors to the conventional s-wave instability, which presumably is inherited from the pairing mechanism of the crystal in bulk form. Note that the mixing discussed here is different than the singlet-triplet mixing that naturally occurs due to the presence of Ising SOC in the absence of external fields (see Supplementary Section 8). In our setup, two strong candidates exist for such an external symmetry-breaking field. Uniaxial strain mixes the s-wave (A$_1$') and d-wave (E') gaps[38,39], giving rise to an overall two-fold anisotropic gap. This is illustrated in Figure 4b, which shows the mixed gap along the $\Gamma$ Fermi surface shown in Figure 1c (details in the Supplementary Section 8). Its two-fold anisotropy contrasts with the six-fold anisotropy of the s-wave gap shown in Figure 4a. Experimentally, residual uniaxial strain could arise for instance due to the exfoliation or encapsulation processes. The applied in-plane magnetic field itself also mixes different gap states, as was shown in Ref.23: due to the presence of the Ising SOC, it mixes the singlet component of the s-wave (A$_1$') gap with the triplet component of the p-wave (E'') gap, giving rise to a two-fold symmetric gap, illustrated in Figure 4c. We found that the two-fold anisotropy of the s+p mixed state is generally weaker than in the s+d mixed state, which can be seen here by comparing Figures 4b and 4c.

Insets in Figure 4 provide cartoon pictures of the possible s+p ($A_1'+E''$) and s+d ($A_1'+E'$) mixed states and their underlying symmetry-breaking fields. While a microscopic calculation connecting the anisotropic magneto-resistance, critical field, and tunneling data presented here with a specific form of the gap function is beyond the scope of this work, we expect that the two-fold anisotropy of the mixed gap function – and of the corresponding spectrum of superconducting fluctuations[40] – will be generally manifested as two-fold anisotropic superconducting properties.

Regardless of which scenario is realized here – spontaneous nematic superconductivity or strong gap-mixing triggered by a small symmetry-breaking field – our results point to a significant contribution of an unconventional pairing mechanism to the superconducting state of few-layer $NbSe_2$. This raises fundamental questions about the origin of such pairing interactions, and opens up fascinating prospects of combining them with non-trivial topological properties that have been predicted in the regime of high magnetic fields. Overall, our work reveals that few-layer TMDs provide a promising framework to realize and explore unconventional superconductivity.

**Methods:**

**Device fabrication**

For the magneto-transport and critical field measurement devices, Cr/PdAu (1nm/7nm) bottom contacts were deposited on top of an atomically-clean piece of hexagonal boron nitride (h-BN) of thickness 20-80nm, above which a piece of 2H-$NbSe_2$ (commercial, from HQ Graphene) and h-BN (20-80nm) was subsequently transferred using the standard dry-transfer technique[37]. The $NbSe_2$ layer number was determined through optical contrast[41] calibrated by atomic-force microscopy. The $NbSe_2$ exfoliation and subsequent hBN-encapsulation is performed in an Ar-filled glovebox to minimize sample degradation. After completing the assembly process, the van der Waals (vdW) heterostructure was taken out of the glove box and etched using reactive ion etching with $O_2$/Ar/$CHF_3$ to open windows on the top h-BN for contacting the pre-patterned bottom contacts. Metal leads of Cr/Pd/Au (1nm/5nm/120nm) were subsequently deposited to connect the bottom contacts to the bonding pads.

For the junction devices, 2H-NbSe$_2$ bulk single crystals were prepared by the iodine-based chemical vapour transport method. CrBr$_3$ and hexagonal boron nitride (h-BN) bulk single crystals were grown by HQ graphene. NbSe$_2$ and CrBr$_3$ were thinned down by the conventional Scotch tape method and assembled by dry transfer technique[42,43]. In detail, both NbSe$_2$ and CrBr$_3$ was first exfoliated on a polydimethylsiloxane (PDMS) polymer substrate. NbSe$_2$ flakes on the PDMS stamp were transferred to a Si/SiO$_2$ (295 nm) substrate which was heated to 70 °C. By searching with an optical microscope, trilayer NbSe$_2$ was identified. A stamp with a thin layer of polypropylene carbonate (PPC) on PDMS was prepared on a glass slide. The trilayer NbSe$_2$ was picked up to the stamp at 45 °C. Thin CrBr$_3$ flakes were identified on the PDMS polymer substrate and aligned to the NbSe$_2$ flake on the stamp. The CrBr$_3$ was picked up from the PDMS substrate to the stamp at 50 °C. The stack was aligned and approached to a pre-patterned metal electrode at 50 °C followed by heating the substrate to 120 °C in order to release the stack. The pre-patterned metal electrodes of Ti (3nm)/Pt (70nm) were prepared by standard photolithography and electron-beam evaporation. The PPC layer on the device was dissolved by anisole. Lastly, the device was single-side encapsulated by h-BN to prevent any degradation in air. All device fabrication procedures occurred in the glove box.

**Measurements**

The magneto-resistance measurements for Device 1 were performed using a Physical Properties Measurement System equipped with a rotating, variable-temperature sample insert and a 9T magnet. A standard lock-in technique was used with an a.c. current of 400 nA (3 µA was used for the R vs. T in Fig 1f only). Device 3 measurements were performed at the National High Magnetic Field Laboratory (NHMFL) in Tallahassee, FL, using a standard lock-in technique in a He$^3$ cryostat equipped with a rotating sample insert and a 35T d.c. magnet. For the measurements at NHMFL, an a.c. current of 1 µA was used.

The junction devices were measured in a separate 9T Physical Property Measurement System at 1.9 K. A rotating probe was used to apply in-plane magnetic field in an angle $\theta$. For the four-probe transport measurement of $NbSe_2$, we chose four electrodes which are in contact with $NbSe_2$ uncovered by $CrBr_3$. A standard lock-in technique was used with an a.c. current of 1 $\mu A$. For the differential conductance measurements, a small a.c. current and d.c. bias current were generated by a lock-in amplifier and a d.c. voltage source meter, respectively, with load resistances in series. The superimposed current flowed through one of the split electrodes of the junction. The a.c. and d.c voltages were measured between the other split electrode and another remote electrode using a pre-amplifier, lock-in amplifier and a voltage meter (see Supplementary Section 6 for measurement schematics). The amplitude of the a.c. voltage was kept below 50 $\mu A$ and d.c. voltage was varied from -8 mV to 8 mV. The differential conductance $G$ was calculated from the ratio of the a.c. current to the measured a.c. voltage. The bias voltage V in the main text corresponds to the measured d.c. voltage.

**Acknowledgements**

We thank Eun-Ah Kim for useful discussions. B.H. and A.H. would like to thank Dave Graf and Scott Maier for their discussions and support related to work done at the National High Magnetic Field Laboratory. Special thanks also to Zhen Jiang for all of the support associated with the Physical Property Measurement System at UMN. The work at UMN was supported primarily by the National Science Foundation (NSF) Materials Research Science and Engineering Center at the University of Minnesota under Award No. DMR-1420013, via an iSuperSeed Award. Portions of the UMN work were conducted in the Minnesota Nano Center, which is supported by the National Science Foundation through the National Nano Coordinated Infrastructure Network (NNCI) under Award Number ECCS-1542202. A portion of this work was performed at the National High Magnetic Field Laboratory, which is supported by National Science Foundation Cooperative Agreement No. DMR-1644779 and the State of Florida. The research at Cornell was supported by the Office of Naval Research (ONR) under award N00014-18-1-2368 for the tunneling measurements, and the National Science Foundation (NSF) under award DMR-1807810 for the fabrication of tunnel junctions. The work in Lausanne was supported by the Swiss National Science Foundation. K.F.M. also acknowledges support from a David and Lucille Packard Fellowship.

**Author contributions**

B.H., A.H., V.S.P. and K.W. designed the magnetoresistance and critical field experiments. B.H. performed the transport measurements at UMN with support from A.H. and K.T. B.H. and A.H. performed the measurements at the NHMFL with support from A.S. B.H. analyzed the data with support from A.H. under the supervision of V.S.P and K.W . A.H., K.T and X.Z. fabricated the magneto-transport heterostructures with support from B.H., under the supervision of K.W. Analytical modeling was performed by D.S., R.M.F. and F.J.B., who also contributed to the interpretation of the results. E.S., J.S. and K.F.M. designed the junction experiments. E.S. and X.X. fabricated and measured the junctions under the supervision of J.S. and K.F.M. E.S. analyzed the junction data under the supervision of J.S. and K.F.M., with input from V.S.P. and R.M.F.  H.B. and L.F. grew the bulk $NbSe_2$ samples for tunnel junction studies. B.H., A.H., E.S., D.S., V.S.P. and R.M.F. co-wrote the manuscript. All authors discussed the results and provided comments on the manuscript.

**Competing interests**

The authors declare no competing interests.

**Additional information**

    **Supplementary information** is available for this paper at [URL inserted by publisher]

**Correspondence and requests for materials** should be addressed to V.S.P. (vpribiag@umn.edu) and K.W. (kewang@umn.edu).

**Data availability**

All data needed to evaluate the conclusions in the paper are available from the corresponding author upon reasonable request.

**Code availability**

All relevant codes needed to evaluate the conclusions in the paper are available from the corresponding author upon reasonable request.

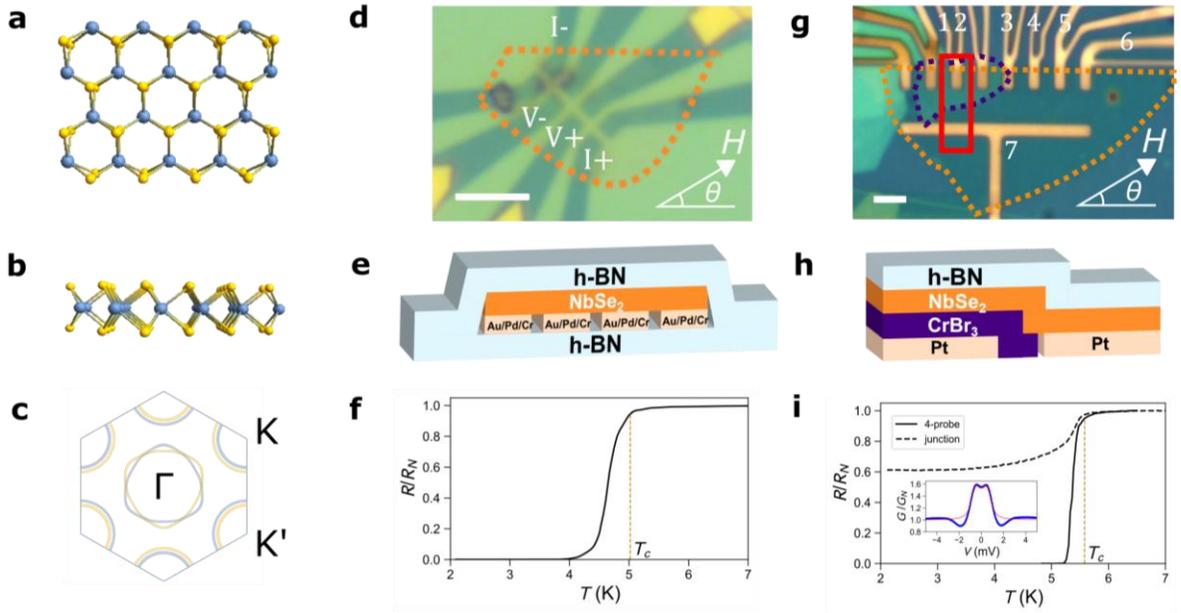

**Fig. 1. Crystal structure, device layout and characterization. a,b.** Top and side views of single-layer 2H-NbSe$_2$ crystal structure, respectively. Odd-layer NbSe$_2$ belongs to the point group D$_{3h}$. **c.** Single-layer NbSe$_2$ Fermi surface in the normal state. The yellow and blue lines represent the spin-split Fermi surface, with each subband corresponding to opposite out-of-plane spin orientations. **d.** Optical image of the 5-layer NbSe$_2$ device (Device 1) used for magnetoresistance measurements (scale bar: 5 μm). Contacts used for magnetoresistance measurements in Figure 2 are indicated. **e.** Diagrammatic cross section of Device 1. **f.** Superconducting transition for Device 1 ($R_N$ = 23.6 Ω) at $\mu_0 H = 0$. $T_c$ is defined as the temperature at which $R = 0.95 R_N$. **g.** Optical image of the 3-layer NbSe$_2$ and bilayer CrBr$_3$ device (Device 2) used for junction studies (scale bar: 5 μm). Contacts 1, 2, 4, and 7 were used in the junction study. Contacts 3, 4, 5, and 6 were used in the four-probe measurement. For Devices 1 and 2, θ is defined as the angle of the magnetic field with respect to the long straight edge of the NbSe$_2$ crystal (see **d** and **g**, orange outline). **h.** Diagrammatic cross section of Device 2. **i.** R vs. T for Device 2, showing junction resistance ($R_N$ = 266 Ω) and 4-probe resistance ($R_N$ = 30.6 Ω) at $\mu_0 H = 0$. The junction resistance has a sharp drop at the superconducting transition. Inset: Differential conductance (blue) and fit to the BTK model (red), indicating a superconducting gap of 0.67 meV, comparable to $T_c$.

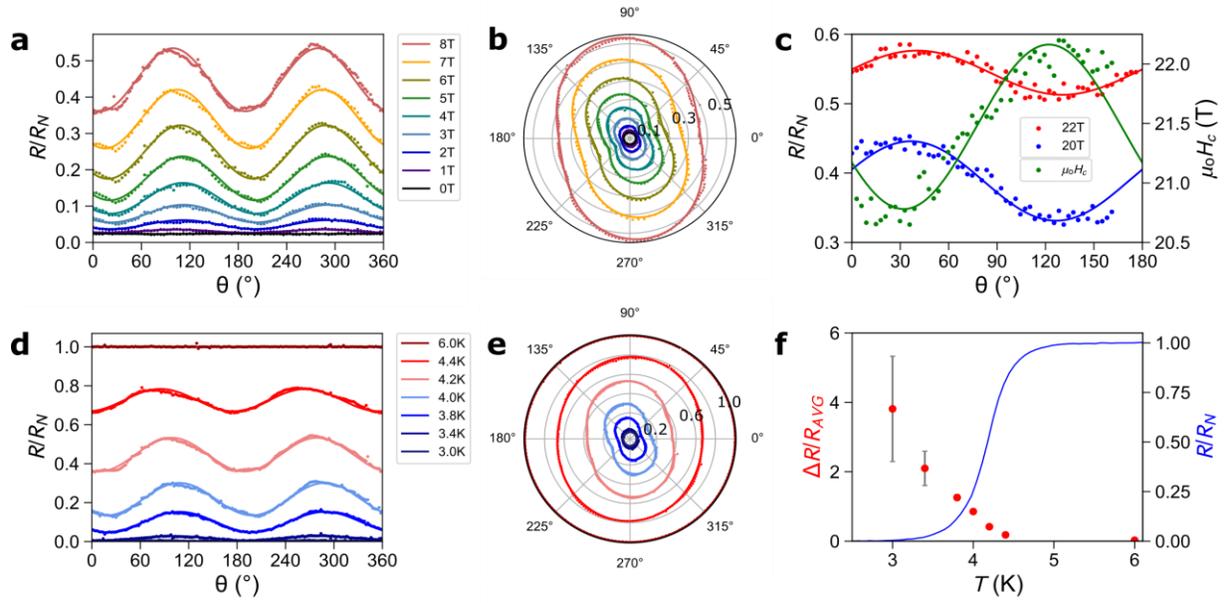

**Fig. 2. Magnetoresistance and critical field signatures of twofold superconducting behavior. a.** Field dependence (T = 4.2K) of magnetoresistance for Device 1. **b**. Radial plot of data in **a**; the radial scale is $R/R_N$. **c.** Critical field (green) and magnetoresistance characterization (blue and red) for Device 3, 5-layer NbSe$_2$ ($R_N$ = 22.2Ω, T = 0.5K). For Device 3, θ is defined with respect to an arbitrary reference. **d.** Temperature dependence ($\mu_0 H = 8T$) of magnetoresistance for Device 1. **e**. Radial plot of data in **d**. Solid lines in **a**-**e** are fits to a $\cos(2\theta + \varphi)$ form. **f.** Relative amplitude ($\Delta R/R_{AVG}$) of the magnetoresistance oscillations shown in **d**; the R vs. T for Device 1 in the presence of $\mu_0 H = 8T$ is plotted in blue. The error bars were calculated using conventional error propagation; the uncertainty of a single data point from **d** was defined as the standard deviation of the 6K data.

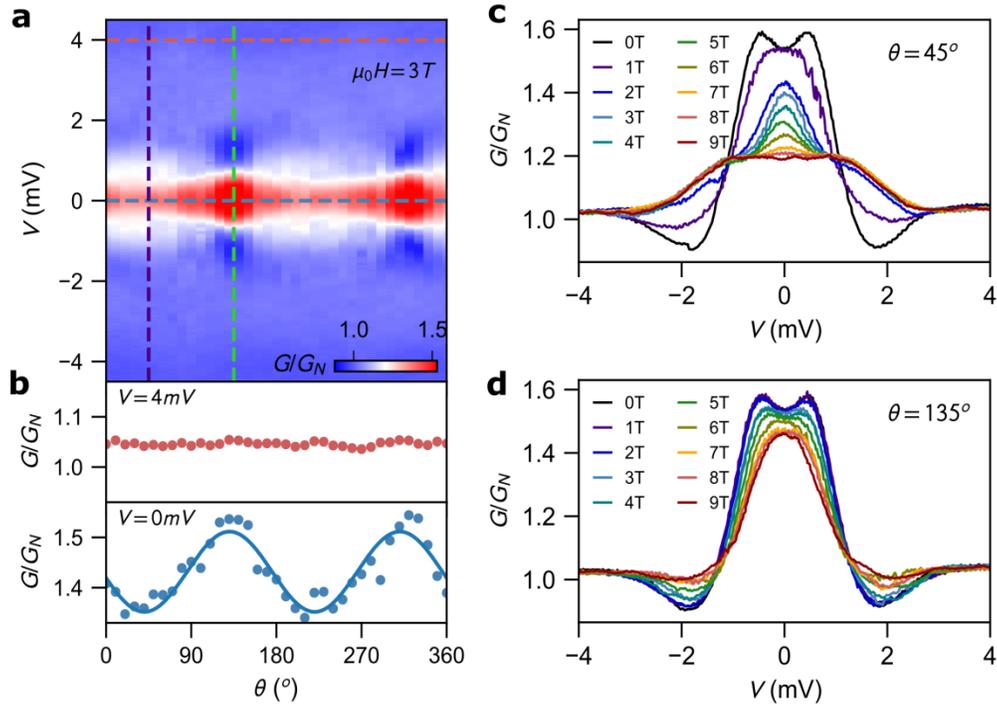

**Fig. 3. Differential conductance spectra under in-plane magnetic field**. **a**. Color plot of the differential conductance of trilayer NbSe$_2$/bilayer CrBr$_3$/Pt junction as a function of the angle θ of the in-plane magnetic field ($\mu_0\text{H} = 3\text{T}$, $\text{T} = 1.9$ K). **b**. $G/G_N$ vs. angle at $V = 0$ and 4 mV (blue and red dashed lines in **a**). The solid line is the best fit to a $\cos(2\theta + \varphi)$ form. **c**, **d**. Field-dependent differential conductance spectra at two fixed in-plane field directions: θ = 45° (**c**, purple dashed line in **a**) and 135° (**d**, green dashed line in **a**).

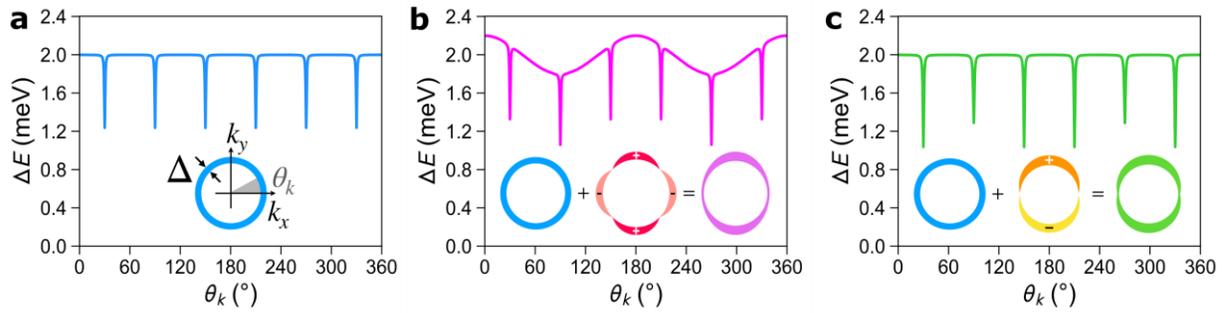

**Fig. 4. Theoretical model for the two-fold anisotropic gap in NbSe$_2$**. All gaps refer to the $\Gamma$ Fermi surfaces shown in Fig. 1c; although two Fermi pockets exist, the gaps are essentially equal in both of them. **a.** A$_1$' (s-wave) gap function. The six-fold anisotropy is a consequence of the Ising-SOC and reflects the three-fold rotational symmetry of the lattice. **b.** Mixing of A$_1$' (s-wave) and E' (d-wave) gaps promoted by uniaxial strain. **c.** Mixing of A$_1$' (s-wave) and E'' (p-wave) gaps promoted by the in-plane magnetic field. Insets: schematics of the superconducting gaps A$_1$' (blue), E' (red), E'' (yellow), A$_1$'+E' (magenta), and A$_1$'+E'' (green).

# Supplementary Information

**S1. Cant angle analysis**

We rule out the possibility that the observed two-fold modulation might have been caused by an out-of-plane component of the applied field, which would have oscillated with the same periodicity due to an unintentional cant of the sample plane with respect to the field rotation plane. The geometry of the canted sample and pertinent parameters are shown in Figure S1.1. The perpendicular field is proportional to $|\sin(\theta + \theta_0)|$, where $\theta_0$ is the angle needed for the field to be totally in the plane of the sample. The voltage response is also proportional to the perpendicular field for moderate $R/R_N$ (Figure S1.2), so the voltage response to an out-of-plane field should be proportional to $|\sin(\theta + \theta_0)|$.

We carried out fits for $\sin[2(\theta + \theta_0)]$ and $|\sin(\theta + \theta_0)|$ to determine the major cause of the sample's response (Figures S1.3 a-d). Moreover, a control test was performed with the same sample, but introducing an intentional (~7°) cant by using a $SiO_x$ chip as shim (Figures S1.3 e-h). We find that $|\sin(\theta + \theta_0)|$ fits the 7° cant data well, and the fit is clearly better than that for $\sin[2(\theta + \theta_0)]$. This demonstrates that magneto-resistance data does take on a $|\sin(\theta + \theta_0)|$ form in the presence of a tilted sample. By comparison, the data with no intentional cant (as in the main text) is fit best by $\sin[2(\theta + \theta_0)]$ and does not fit well to $|\sin(\theta + \theta_0)|$, excluding the possibility that the effects reported in the main text are due to an accidental tilt of the sample with respect to the field plane. This is further supported by our observations in one sample that the angle of minimum resistance of the twofold oscillations was affected by temperature and field modulation (Figure 2), which is inconsistent with the view of these oscillations being caused by misalignment (further covered in S3).

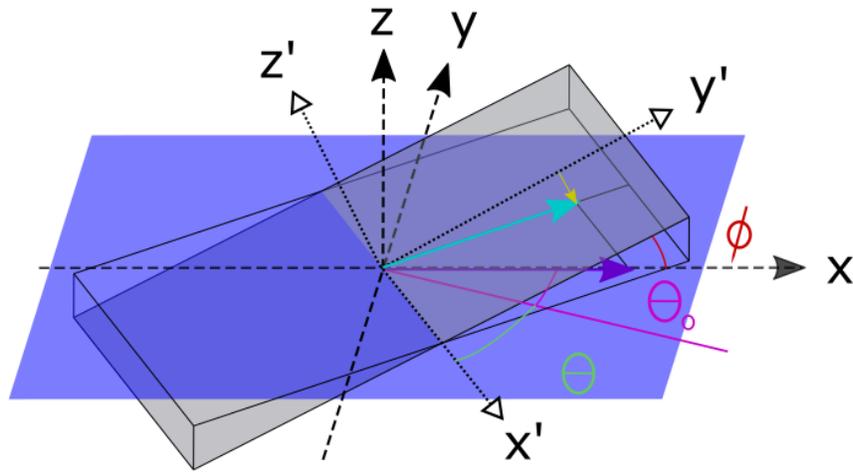

**Figure S1.1. Geometry of a canted sample**. The field lies in the x-y plane, and the sample in the x'-y' plane. The field component perpendicular to the plane of the sample takes the form $\mu_0 H |\sin(\theta + \theta_0)| \sin(\phi)$. Here, $\theta$ is the angle in the x-y plane, $\theta_0$ is the angle at which the field lies entirely in the plane of the sample (x'-y'), and $\phi$ is the sample's cant angle from the plane in which the field rotates (angle between the x-y and x'-y' planes). The field plane is blue, and the sample plane is grey. $\mu_0 H$, $\mu_0 \sin(\theta + \theta_0)$ and $\mu_0 H |\sin(\theta + \theta_0)| \sin(\phi)$ are indicated by purple, light blue and yellow arrows, respectively.

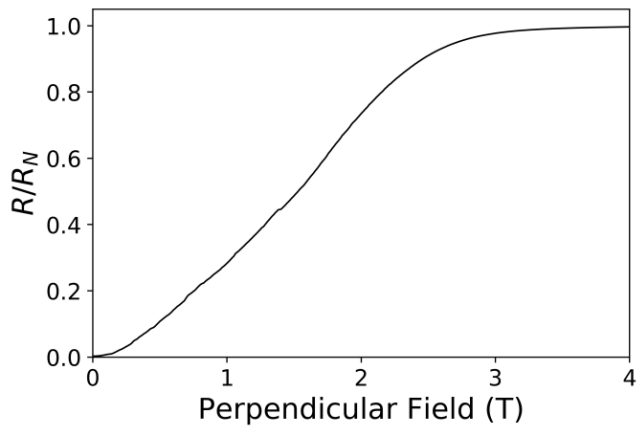

**Figure S1.2. Magnetoresistance characterization for out-of-plane sweep using Device 1**. For a range of moderate $R/R_N$, resistance is roughly linear with perpendicular field (T=1.8K).

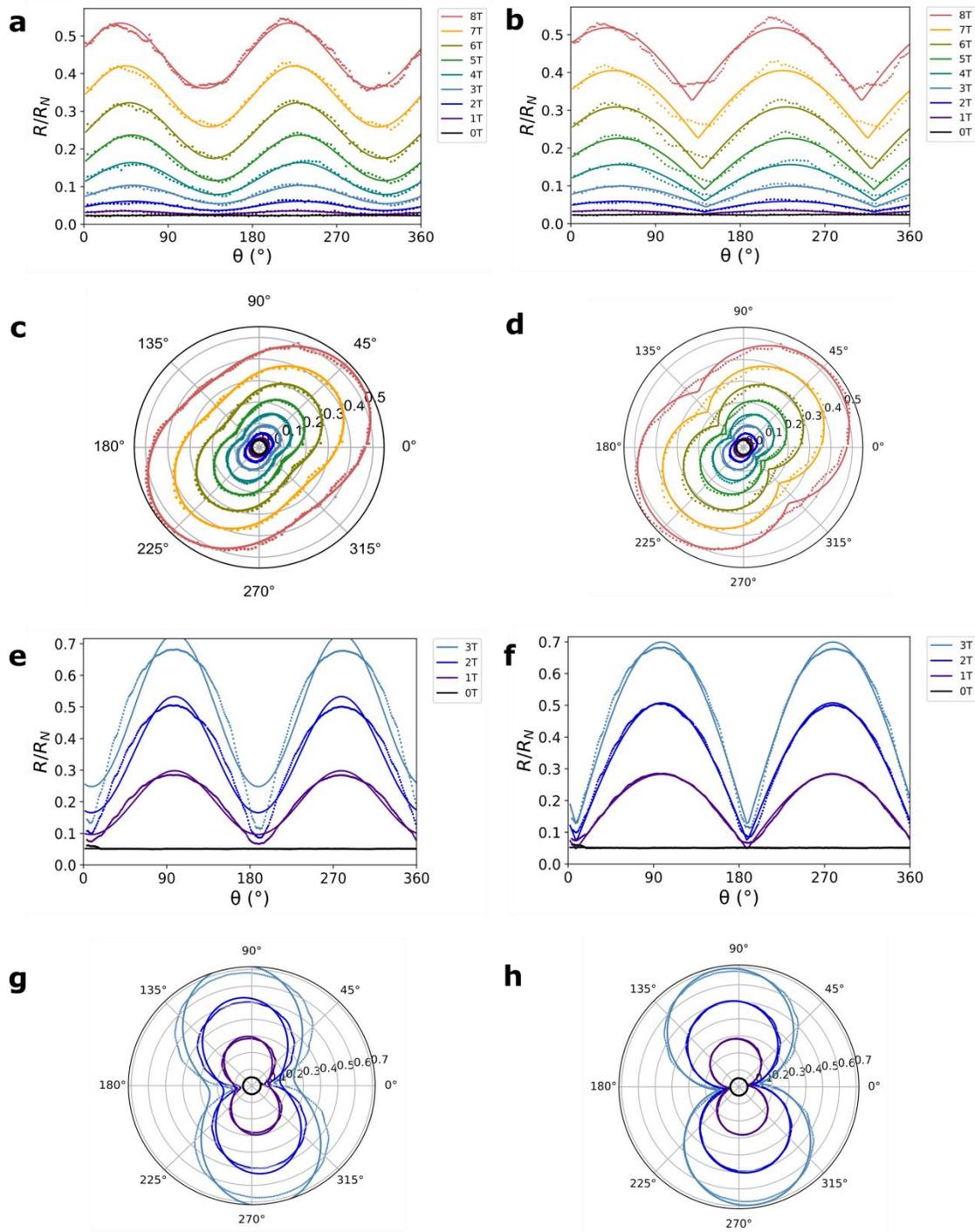

**Figure S1.3. Data are qualitatively different between canted and non-canted samples**. **a-d**: Device 1 with no cant. **e-h.**: Device 1 with intentional cant. **a, c, e, g**: Fits to $\sin[2(\theta + \theta_0)]$. **b, d, f, h**: Fits to $|\sin(\theta + \theta_0)|$. Radial axis for polar plots is $R/R_N$. Fits are solid curves, and experimental data are discrete points.

## S2. Quantitative analysis of magneto-resistance to exclude plane misalignment

The general model for misalignment-induced magnetoresistance oscillations is shown in Figure S2a where, due to misalignment angle ϕ, the normal vector n̂ of the sample precesses in a cone with in-plane rotation. This would lead to resistance maxima at the two angles corresponding to max out-of-plane field, $\theta_{|\vec{H_\perp}|_{max}}$, due to thin superconductors being more vulnerable to out-of-plane fields than in-plane fields due to vortex formation. The hypothetical ϕ was found by first measuring one sample's in-plane magnetoresistance oscillations (Figure S2b) and then directly measuring the transition curve with respect to $H_\perp$ (Figure S2c) after changing the sample holder, re-bonding and confirming an essentially unchanged contact resistance. By examining the voltage fluctuations at the temperature matching that used for the transition curve, one can estimate the associated max out-of-plane field if the oscillations were due to misalignment. Comparing this to the applied field in S2b, one gets an associated misalignment ϕ~2.44°, which is much higher than what is expected to be realistic for our measurement setup. Image analysis showed an estimated ϕ on the order of tenths of a degree, thus being approximately one order of magnitude smaller than that capable of explaining our data. In addition, the hBN-SiO$_2$ and hBN-NbSe$_2$ interfaces are unlikely to host this misalignment due to the associated nm-scale topographic variation and μm length scale of the sample.

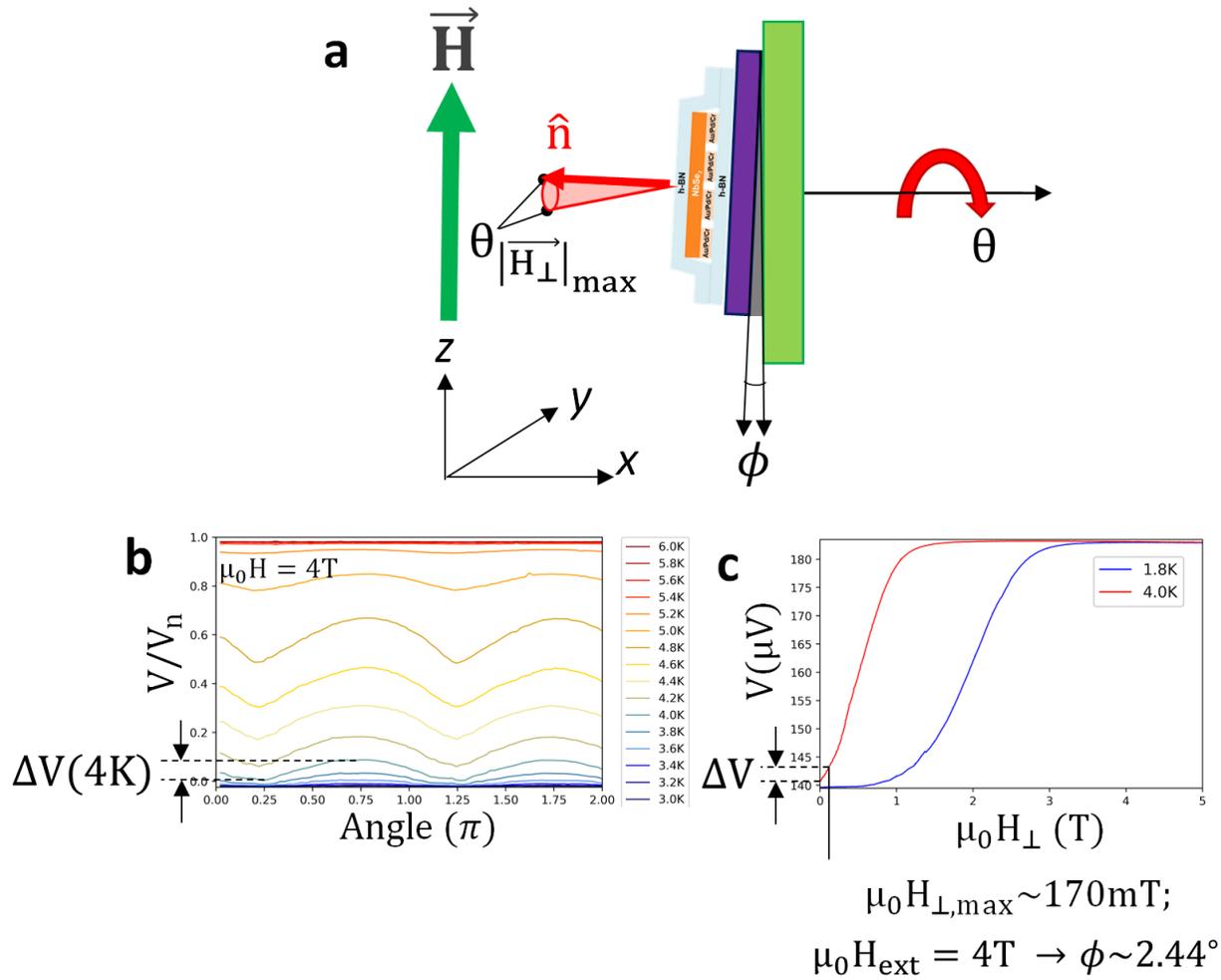

**Figure S2. Quantitative analysis of magneto-resistance to exclude plane misalignment. a.** Schematic showing geometry of setup with sample cant angle of ϕ. **b.** Temperature-dependent data from Device 1 in an in-plane 4T field. **c.** Magnetoresistance of out-of-plane field sweep (three-point measurement using an a.c. current of $1\mu A$).

### S3. Shifting minima

Another piece of evidence against an out-of-plane effect was a shifting minimum angle in the data. For a purely out-of-plane effect, no shift should be expected – this is confirmed in Figure S3 for the canted Device 1, whose minima had a range of 4° (the resolution of the measurement). Instead, Device 1 without a cant showed a minimum range of 16° for temperature dependence with a general trend to the left, and the field dependence shows a range of 32°. Notable is that this phase shift with both temperature and field modulation implies a dependence on location along the superconducting transition, rather than being due to extrinsic effects.

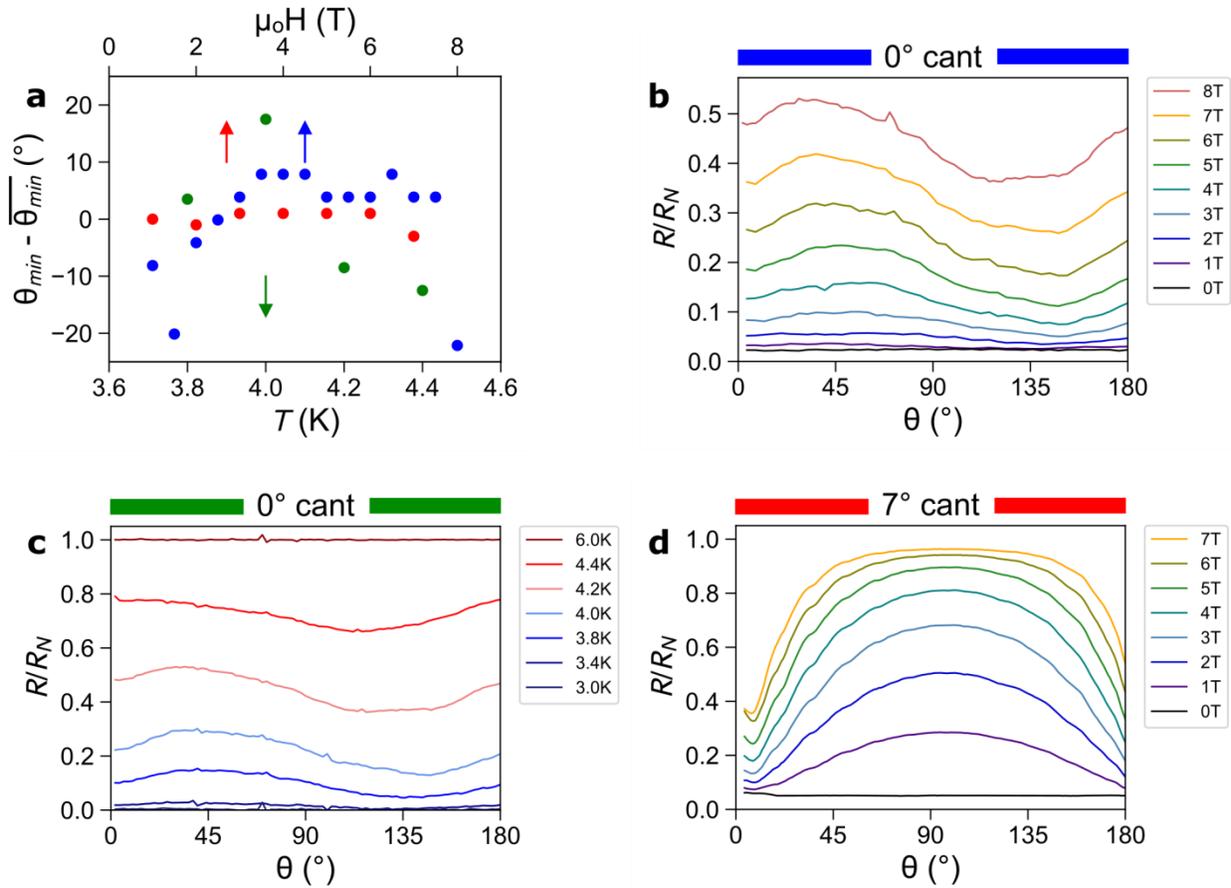

**Figure S3. Magnetoresistance minimum shift present for 0° cant, absent for 7° cant. a**. Angle at which minimum of magnetoresistance occurs for the three cases shown in panels **b**-**d**, centered around each set's average minimum angle. **b.** Field dependence of Device 1 with 0° cant (average minimum: 144°). **c.** Temperature dependence of Device 1 with 0° cant (average minimum: 133°). **d.** Field dependence of Device 1 with 7° cant (average minimum: 10°). The minimum for the 7° cant shifts very little (almost not at all) compared to that of the 0° cant. Note: the data for the 0° cant is the raw data of that shown in the main text.

## S4. Characterization of Device 3.

Device 3 was made from a 5-layer NbSe$_2$ sample, the same as Device 1. The R vs. T (Figure S4a) for this sample showed an $R_N$ of 20.4$\Omega$ and $T_c \approx$6.2K. This gives a superconducting gap $\Delta \approx 1.76 k_B T_c \approx 0.94$ meV. Although the transition width is wider than in Device 1 and less uniform at low resistances, the magnetoresistance and critical field data were taken at a smooth part of the transition ($\sim 0.35 R_N$ to $\sim 0.6 R_N$) and showed no anomalies.

Resistance dependence on perpendicular and parallel fields were also studied at 0.5K (Figure S4b). The out-of-plane critical field was ~4T, and the in-plane critical field was ~27T. These two measurements were taken on the same holder, and immediately following one another, but with the holder turned 90° from the first to second measurement such that the rotation in-plane with the field became a rotation out-of-plane.

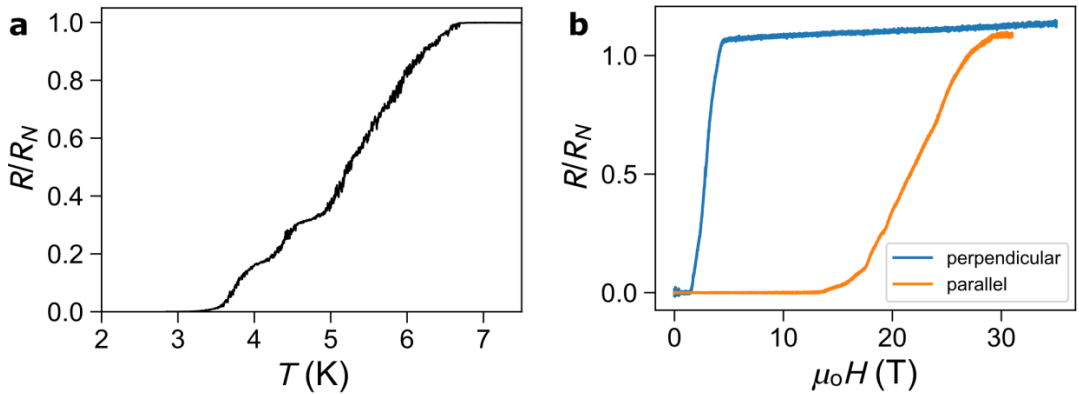

**Figure S4. Characterization for Device 3**. **a.** Normalized R vs. T. **b.** Normalized R vs. magnetic field for both when the field and sample plane are parallel and perpendicular. $R_N = 20.4\Omega$.

**S5. Directional Analysis of Magnetotransport across multiple devices.**

Directions of current, voltage drop, and that of the magnetic field associated with resistance extrema are shown in Figures S5a-c, with additional images of the corresponding NbSe$_2$ flakes as guides to the eye when needed. It can be seen that the direction of the current and voltage drop show no correspondence with the field direction of minimum resistance $\hat{H}_{R\,min}$. However, examination of $\hat{H}_{R\,min}$ across all samples shows a repeated alignment with the long, straight edge of the NbSe$_2$ flakes. Figure S5d shows the associated magnetoresistance oscillations used for each device to determine the orientation of $\hat{H}_{R_{min}}$.

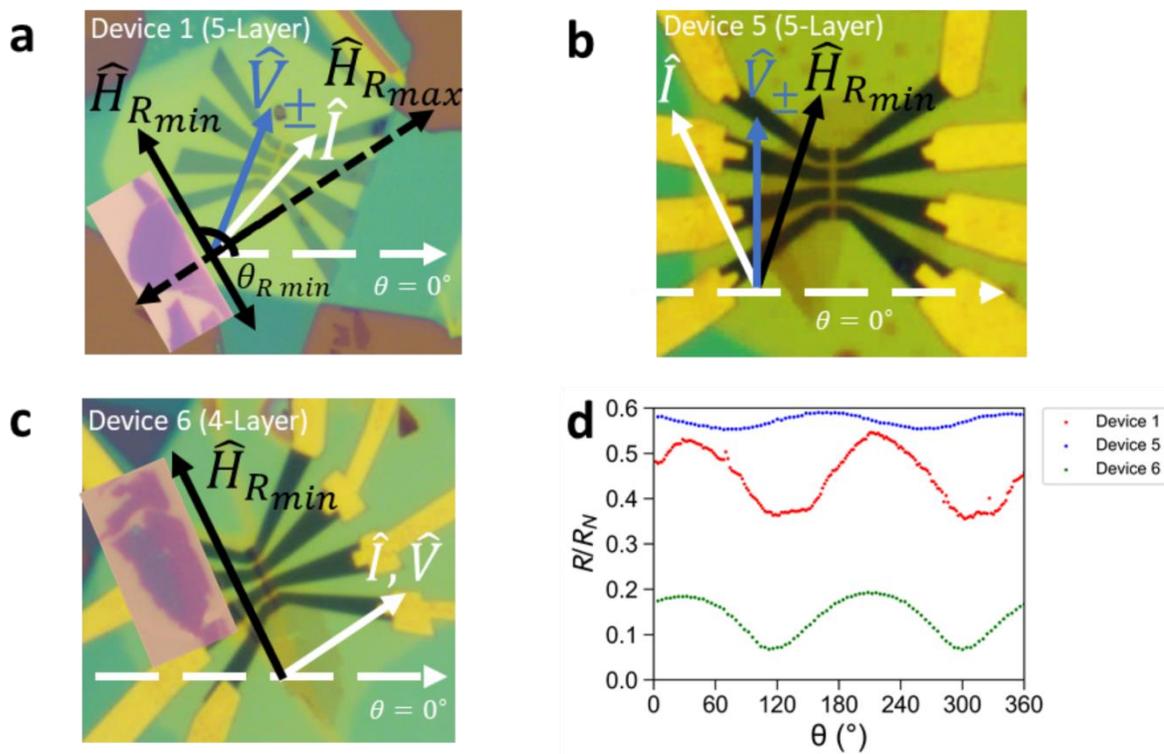

**Figure S5. Directional analysis of magnetotransport across several devices. a.-c.** Device images of Devices 1, 5, and 6 along with the associated direction of the field of minimum resistance as well as the current and voltage directions. Pictures of the NbSe$_2$ flakes are provided as guides to the eye as needed. In Device 1, the associated field of maximum resistance is provided as well as an example. **d.** Plot of the magnetoresistance oscillations used for each device for determination of $\hat{H}_{R_{min}}$. The angle of minimum resistance for Devices 1, 5, and 6 correspond to 122°, 68°, and 120° respectively.

## S6. Differential conductance measurement schematics

Figure S6 shows the setup for measuring the differential conductance of the junction highlighted by a red square. We use a lock-in amplifier and voltage source meter as the a.c. and d.c. voltage source, respectively. Each instrument is connected to a load resistance to convert the voltage signal to a current signal. A typical d.c. and a.c. load resistance is 100 kΩ and 1MΩ, respectively. The superimposed a.c. and d.c. current is fed to one of the split electrodes. The a.c. and d.c. voltage signal is measured between the other split electrode and another remote electrode. The differential conductance is defined as the ratio of the applied ac current to the measured ac voltage. We normalize the differential conductance $G$ to the zero bias differential conductance in the normal metal state $G_N$ measured at $T \approx$ 5.8 K. All spectra reported are plotted in terms of $G/G_N$ vs. the measured d.c. bias voltage.

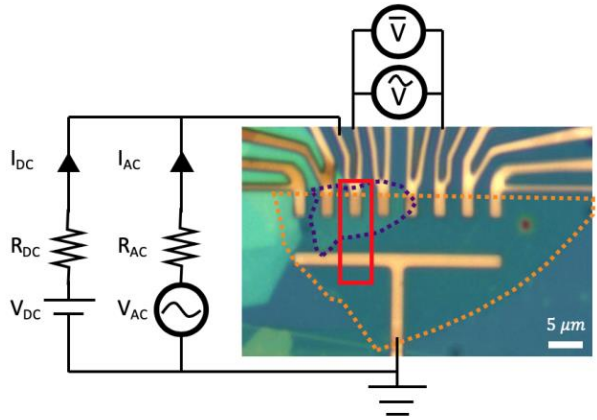

**Figure S6. Schematic of differential conductance measurement.**

## S7. Differential conductance measurements on Device 4

We present a second device which is constructed with a magnetic metal electrode (multilayer Co/Pt) which shows identical results as the one introduced in the main text. Pre-patterned magnetic metal electrodes are prepared by standard photolithography on a Si/SiO$_2$ (290 nm) substrate and e-beam evaporation of 3 nm Ti/10 nm Pt/(0.3 nm Co/1 nm Pt)×8/0.5 nm Co/2.5 nm Au/0.4 nm AlO$_x$. Multilayer Co/Pt is a ferromagnetic metal with an easy-axis in the out-of-plane direction [S1]. The measured saturation field for the out-of-plane and in-plane direction is ~0.5 T and ~0.8T, respectively. The 2.5 nm Au prevents oxidation of Co and the 0.4 nm AlO$_x$ layer serves as an insulating barrier. A trilayer NbSe$_2$ flake and a h-BN protection layer is transferred in sequence on to the prepatterned substrate by the standard dry transfer technique (Figure S7a). The normal state resistance of the junction is $R_N$ = 11 Ω which is more than an order of magnitude smaller than the CrBr$_3$ barrier device.

We repeat the measurement as introduced in the main text. An in-plane magnetic field of 3T is applied which saturates magnetic moment of the Co/Pt electrodes in the in-plane direction. Differential conductance spectra are acquired as the magnetic field is rotated in the in-plane direction. The results are summarized in Figure S7b. Consistent to the main text, there is a two-fold dependency of the spectra in terms of the angle θ. The zero-bias normalized conductance G/G$_N$ vs. angle θ shows a cos(2θ) dependence while at larger bias the dependency is weak. When the in-plane field is fixed in θ = 90°, the differential conductance within the gap suppresses quickly in larger field, similar to Figure 3c in the main text. On the other hand, when the field is directed in θ = 160°, the tunneling spectra is more robust showing little suppression to the gap like that observed in Figure 3d. We find that tunneling to NbSe$_2$ either through a magnetic insulating barrier or from a magnetic electrode results in a two-fold modulation in the tunneling spectra.

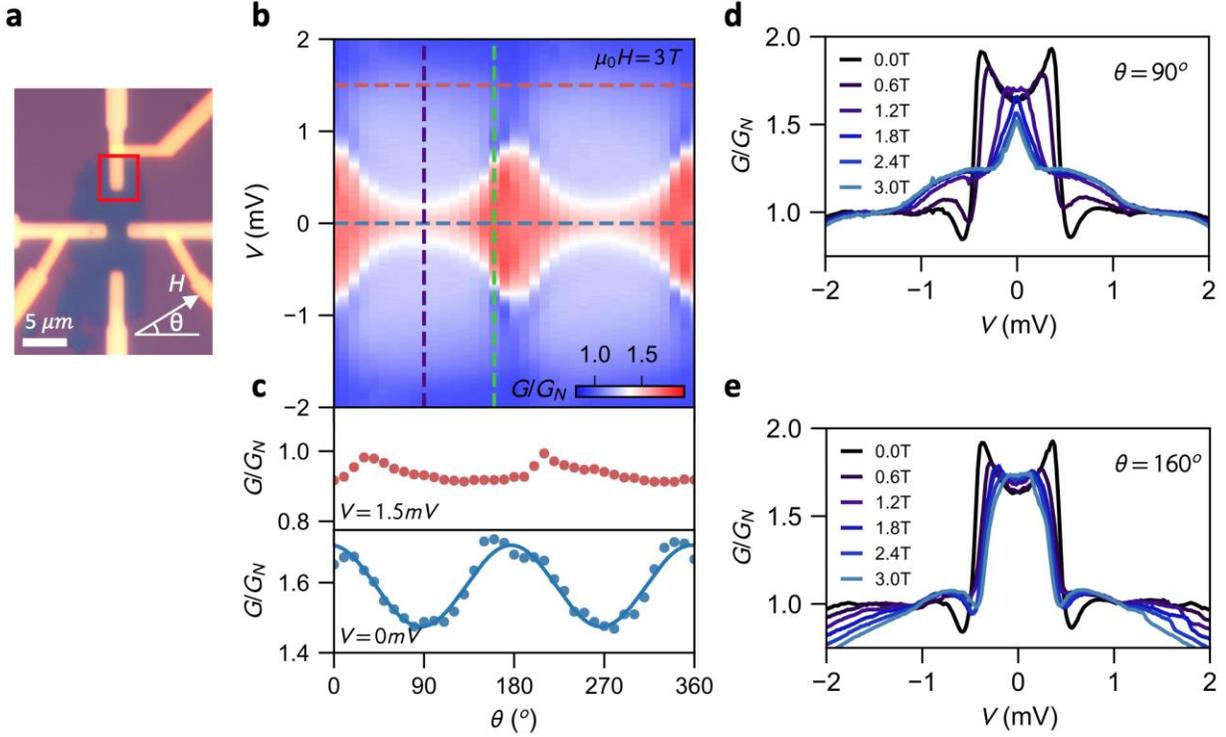

**Figure S7. Differential conductance measurement on Device 4. a**. Optical image of Device 4. The visible prepatterned electrodes are multilayers of Co/Pt. The blue flake is a trilayer NbSe$_2$ and the covering h-BN is not shown in this image. The in-plane magnetic field angle θ is defined as illustrated. **b.** Contour plot of the differential conductance spectra under in-plane magnetic field at 2.3 K. **c.** $G/G_N$ vs. angle at $V = 0$ and 1.5 mV (blue and red dashed lines in **b**). The solid line is the best fit to a cos(2θ) form. **d, e**. Field-dependent differential conductance spectra at two fixed in-plane field directions: θ = 90° (**d**, purple dashed line in **b**) and 160° (**e**, green dashed line in **b**).

## S8. Phenomenological Description of Superconducting Anisotropy in NbSe$_2$

Since we expect the behavior of few-layer NbSe$_2$ to be representative of the monolayer compound, as long as the inter-layer coupling is weak, we consider the point group of the latter, which is $D_{3h}$. This point group is non-centrosymmetric, reflecting the fact that it incorporates the effects of the Ising spin-orbit coupling (SOC).

The possible superconducting gaps, which we denote $\widetilde{\Delta}_{\Gamma,m}^{(\alpha)}$, can be classified based on the irreducible representation Γ of $D_{3h}$ under which they transform, as well as whether they correspond to spin-singlet ($\alpha = s$) or to spin-triplet ($\alpha = t$) Cooper pairs. When Γ is not 1-dimensional, we use the index $m$ to distinguish different components of the corresponding multiplet. We note that since $D_{3h}$ incorporates Ising SOC, a given $\Gamma$ contains both singlet and triplet gap components, and the solutions of the superconducting gap equation are a mixture of spin-singlet and spin-triplet Cooper pairs[12,14,24]

| $\Gamma$ | $m$ | $\Theta^{(s)}_{\Gamma,m}(\mathbf{p})$ | $\Theta^{(t)}_{\Gamma,m}(\mathbf{p})$ |
|---|---|---|---|
| $A'_1$ | 1 | $i\sigma^y$ | $\cos3\theta\ \sigma^z i\sigma^y$ |
| $E'$ | 1 | $\cos2\theta\ i\sigma^y$ | $\cos\theta\ \sigma^z i\sigma^y$ |
|  | 2 | $\sin2\theta\ i\sigma^y$ | $\sin\theta\ \sigma^z i\sigma^y$ |
| $E''$ | 1 |  | $\cos3\theta\ \sigma^x i\sigma^y$,  $(\cos\theta\ \sigma^x - \sin\theta\ \sigma^y)i\sigma^y$ |
|  | 2 |  | $\cos3\theta\ \sigma^y i\sigma^y$,  $(\sin\theta\ \sigma^x + \cos\theta\ \sigma^y)i\sigma^y$ |

**Table S8.1. Irreducible representations of $D_{3h}$ relevant to our analysis.** We exclude irreps that vanish in the monolayer ($p_z = 0$) limit. The singlet component of $A_1'$ $E'$ can be associated to an $s$-wave ($d$-wave) gap, while the lowest harmonic in the triplet $E''$ irrep can be associated with a $p$-wave order parameter. Here we take one of the $K$ points to lie along the $p_x$ axis, i.e. $\theta = 0$.

Here we focus on the $\Gamma$ pocket (not to be confused with the irrep index), since we expect the gap associated with $\pm K$ pockets to be isotropic[14]. The corresponding superconducting gap $\widetilde{\Delta}(\mathbf{p})$ can then be expressed as:

$$\widetilde{\Delta}(\mathbf{p}) = \sum_{\Gamma\alpha m} \Theta^{(\alpha)}_{\Gamma,m}(\mathbf{p})\Delta^{(\alpha)}_{\Gamma,m} \tag{1}$$

where $\Delta^{(\alpha)}_{\Gamma,m}$ is a complex scalar parameterizing the magnitude of the contribution to the superconducting gap from the irrep $\Gamma$, in pairing channel $\alpha$. We take $\Theta^{(\alpha)}_{\Gamma,m}(\mathbf{p})$ to be functions of only the angle $\theta$ with respect to the Brillouin zone center. Note that by particle hole symmetry, $\widetilde{\Delta}(\mathbf{p}) = -\widetilde{\Delta}^T(-\mathbf{p})$, which implies that $\Theta^{(s)}_{\Gamma,m}(\mathbf{p})$ ($\Theta^{(t)}_{\Gamma,m}(\mathbf{p})$) is an even (odd) function of $\mathbf{p}$.

The irreps relevant to our analysis here are shown in Table S8.1. To make a connection with the usual classification of gaps in isotropic space, we note that these are an $s$-wave singlet gap and $f$-wave triplet gap in the $A_1'$ irrep; a $d$-wave singlet and a $p$-wave triplet in the $E'$ irrep; and $p$-or $f$-wave triplets in the $E''$ irrep. We have omitted most irreps corresponding to gaps that are non-uniform on the $K$ pockets, which do not contribute to the leading-order superconducting instabilities.[14] The exception is the E' irrep, which we keep because it couples to A$_1$' by strain, and the $p$-wave gap in the E'' irrep, which can couple to the magnetic field, as these can generate a two-fold anisotropy in the superconducting gap. We have also kept only irreps relevant to the monolayer limit (i.e. independent of the $\hat{z}$ direction).

A few remarks are in order. First, because of the Ising spin-orbit coupling, the A$_1$' and E' irreps contain both singlet and triplet gaps. The E'' irrep, however, contains only triplet components as the corresponding singlet gap cannot be realized in a $\hat{z}$-independent way. Second, the E' and E'' irreps are 2-dimensional, leading to two degenerate solutions to the gap equation.

In order to determine the form of the superconducting gap in the presence of strain and a magnetic field, we use a phenomenological approach, analyzing what terms quadratic in the superconducting gaps are allowed in the Ginzburg-Landau free energy. Within this approach the functional form of the gap is not determined, only the relative weights of each irrep. In practice, one of the terms within a given irrep in Table S8.1 is usually dominant, and the rest can be dropped (which term is relevant is not determined within this approach). Here we focus on terms involving the singlet component of the gap in the A$_1$' ($s$-wave) channel, which we expect to be large compared to its triplet counterpart, and similarly assume that only one of any of the terms in the E' and E'' irreps is relevant. We focus on the singlet $d$-wave term in E' and the triplet $p$-wave term in E'' as those give rise to the two-fold anisotropy in the spectrum. We can therefore drop the superscript on the coefficients in front of the gap function within a given irrep and label them as $\Delta_{A_1'}$, $\mathbf{\Delta}_{E'}$, and $\mathbf{\Delta}_{E''}$, where the latter are 2-component vectors representing the doublet.

We assume that the dominant superconducting instability is in the A$_1$' irrep, which contains singlet $s$-wave terms. Near the superconducting transition temperature (in the absence of a magnetic field or strain), the free energy has the form

$$\mathcal{F}_0 = \mathcal{F}_0[\Delta_{A_1'}] + \tfrac{1}{2}(\chi_{E'}^{-1}(T)|\mathbf{\Delta}_{E'}|^2 + \chi_{E''}^{-1}(T)|\mathbf{\Delta}_{E''}|^2) \ . \tag{2}$$

Here $\chi_\Gamma(T) \propto 1/(T-T_{c,\Gamma})$ is the susceptibility associated with superconducting fluctuations in the $\Gamma$ irrep, with $T_{c,\Gamma}$ being the corresponding transition temperature. We leave here the precise form of $\mathcal{F}_0[\Delta_{A_1'}]$ unspecified. Our main assumption is that the leading superconducting instability takes place at a temperature $T_{c,A_1'}$ that is larger than $T_{c,E'}$ and $T_{c,E''}$.

In the presence of perturbations such as strain and magnetic field, which do not respect the $D_{3h}$ symmetry, additional terms in the free energy that mix the different irreducible representations at quadratic order in the $\Delta_{\Gamma,m}$ are allowed. We consider two mixing terms that can lead to a twofold anisotropy in the superconducting gap.

First, since the in-plane magnetic field $\mathbf{b}$ (in units of the Bohr magneton times the Landé g-factor, $\mathbf{b} = g_L \mu_B \mathbf{B}$) transforms according to the E″ irrep and the Ising spin-orbit coupling vector $\lambda p_F^3 \cos 3\theta \hat{\mathbf{z}}$ transforms according to $A_1'$, a singlet $A_1'$-triplet E″ mixing term of the form $\lambda p_F^3 \Delta_{A_1'} \mathbf{b} \cdot (\hat{z} \times \boldsymbol{\Delta}_{E''})$ is allowed, consistent with the results of Ref. 23. The resulting term in the free energy is

$$\mathcal{F}[\Delta_{A_1'}, \boldsymbol{\Delta}_{E''}] = \lambda_1 \mathrm{Re}\left[\Delta_{A_1'}^*\left(b_y \Delta_{E'',1} - b_x \Delta_{E'',2}\right)\right] . \tag{3}$$

where $\lambda_1$ is some coupling constant proportional to $\lambda$. Minimizing the resulting free energy with respect to $\Delta_{E'',m}$, we find that

$$(\Delta_{E'',1}, \Delta_{E'',2}) = -\lambda_1 \chi_{E''} \Delta_{A_1'}(b_y, -b_x) \tag{4}$$

and the total superconducting gap is

$$\widetilde{\Delta}(\mathbf{p}) = \Delta_{A_1'}\left[\Theta(\mathbf{p}) - \lambda_1 \chi_{E''} \boldsymbol{\Theta}_{E''}(\mathbf{p}) \cdot (\hat{z} \times \mathbf{b})\right] \tag{5}$$

which is a mixture of the s-wave singlet $A_1'$ gap and the p-wave triplet E″ gap. The magnitude of $\Delta_{E''}$ (relative to that of $\Delta_{A_1'}$) is set by the coefficient $\lambda_1 \chi_{E''} |b|$. Since the magnetic field $b$ is a small perturbation, generically we expect this coefficient to be small, and the corresponding mixing to be weak. However, if $T_{c,E''}$ is comparable to $T_{c,A_1'}$, then in the vicinity of $T_{c,A_1'}$ the susceptibility $\chi_{E''}(T) \sim 1/(T_{c,A_1'} - T_{c,E''})$ is very large. Thus we expect that weak magnetic fields can lead to appreciable mixing between the s-wave singlet and the p-wave triplet only when $T_{c,A_1'}/T_{c,E''}$ is close to 1 – i.e. only when these two superconducting instabilities are close competitors.

Second, the strain combination $\boldsymbol{\mathcal{E}} = (\varepsilon_{xx} - \varepsilon_{yy}, 2\varepsilon_{xy})$ transforms according to the E′ irreducible representation. Here, $\varepsilon_{ij} \equiv \frac{1}{2}(\partial_i u_j + \partial_j u_i)$ is the strain tensor and $\mathbf{u}$ is the displacement vector. This leads to a coupling between the singlet components of the $A_1'$ and E′ gaps of the form:

$$\mathcal{F}[\Delta_{A'_1}, \boldsymbol{\Delta}_{E''}] = \lambda_2\left[(\varepsilon_{xx} - \varepsilon_{yy})\mathrm{Re}\left(\Delta_{A_1'}^* \Delta_{E',1}\right) + 2\varepsilon_{xy}\mathrm{Re}\left(\Delta_{A_1'}^* \Delta_{E',2}\right)\right] . \tag{6}$$

where $\lambda_2$ is some coupling constant. Minimizing the free energy leads to a mixed $s$-wave and $d$-wave superconducting gap

$$\tilde{\Delta}(\mathbf{p}) = \Delta_{A_1'}[\Theta_{A_1'}(\mathbf{p}) - \lambda_2 \chi_{E'} \Theta_{E'} \cdot \mathcal{E}] \tag{7}$$

Again, small strains will only lead to a significant E' component in the superconducting gap if the $A_1'$ and E' superconducting instabilities have similar transition temperatures.

Fig. 4 of the main text shows the resulting 2-fold anisotropy in the superconducting gap around the Γ pocket. We obtained the plots using a simple model of the single-body Hamiltonian in momentum space:

$$H(\mathbf{p}) = \sum_{\mathbf{p}s} \varepsilon(\mathbf{p}) c_{\mathbf{p}s}^\dagger c_{\mathbf{p}s} + \sum_{\mathbf{p}ss'} c_{\mathbf{p}s}^\dagger [\lambda p_F^3 \cos 3\theta \sigma^z + \mathbf{b} \cdot \boldsymbol{\sigma}]_{ss'} c_{\mathbf{p}s'} = \sum_{\mathbf{p}ss'} c_{\mathbf{p}s}^\dagger [\mathcal{H}_0]_{ss'} c_{\mathbf{p}s'} \tag{8}$$

where $s$ and $s'$ are spin indices,

$$\varepsilon(\mathbf{p}) = -\frac{p^2}{2m} - \mu, \tag{9}$$

and $\theta$, as above, is the angle measured from the center of the Γ pocket with $\theta = 0$ corresponding to the direction towards the $K$ point. We use $\lambda p_F^3 = 40$ meV, which is the Ising SOC value reported in the literature once averaged over all Fermi surfaces [22]. We also set $\mu = 0.4$ eV, which gives $m = p_F^2/2\mu \approx 2m_0$ when the wavenumber corresponding to the Fermi momentum is $p_F/\hbar = 0.45$ Å$^{-1}$ [S2]. Here, $m_0$ is the electron rest mass. All these parameters were chosen so that the Fermi surface of the toy-model parabolic dispersion is similar to that obtained from first principle calculations[12,22,24]. These values should thus not be understood as actual parameters that a first-principle calculation would yield. The key point is that our conclusions do not rely on this set of parameter values, as they are used to illustrate the two-fold gap anisotropy that must follow from mixing an $A_1'$ gap with either E' or E'' gaps. Finally, for the magnetic field, we use $b = 1$ meV, which is about 8.3 T.

The superconducting excitation spectrum is then given by the eigenvalues of the Bogolyubov-de Gennes (BdG) Hamiltonian:

$$H_{BdG} = \frac{1}{2} \sum_{\mathbf{p}} \Psi_{\mathbf{p}s}^\dagger \mathcal{H}(\mathbf{p}) \Psi_{\mathbf{p}s'} \tag{10}$$

where we use the Nambu-Gor'kov representation $\Psi_{\mathbf{p}s} = (c_{\mathbf{p}s}, c_{-\mathbf{p}s}^\dagger)^T$ and

$$\mathcal{H}(\mathbf{p}) = \begin{pmatrix} \mathcal{H}_0(\mathbf{p}) & \tilde{\Delta}(\mathbf{p}) \\ \tilde{\Delta}^\dagger(\mathbf{p}) & -\mathcal{H}_0^T(-\mathbf{p}) \end{pmatrix}. \tag{11}$$

We took $\widetilde{\Delta}(\mathbf{p})$ as in Eqs. (5) and (7). We took the singlet $d$-wave term for E' and the triplet $p$-wave term for E'' from Table 1. For all plots, we took $\Delta_{A_1'} = 2$ meV. For the mixed $A_1'$/E' gaps, we took $|\Delta_{E'}| = 0.2$ meV and an uniaxial strain along the $x$ axis, $\mathcal{E} \propto (1,0)$. For the mixed $A_1'$/E'' gap, the mixing is relatively weak unless the mixed gaps are close in magnitude; for this reason, we used $|\Delta_{E''}| = 1$ meV in the plot. Note that we plot the energy gap in the spectrum of $H_{BdG}$ (i.e. difference between bottom positive energy band and top negative energy band) along both the inner and outer Fermi surfaces. The resulting gaps on the inner and outer surfaces are identical within the resolution of the figure.

A second potential source of anisotropy comes from the spatial variations of the doublet gaps in the E' and E'' irreps, i.e. from the gradient terms of the Ginzburg-Landau free-energy. For the one-component $A_1'$ irrep, the only allowed gradient term is of the form $\mathcal{F}_\nabla^{(A_1')} = K_0 |\nabla D^{(0)}|^2$, which is clearly rotationally invariant. For the doublet E' and E'' representations, however, the possible terms in $\mathcal{F}$ involving gradients of the gap functions are [S3].

$$\mathcal{F}_\nabla^{(\Gamma)} = K_1 |\nabla \cdot \mathbf{\Delta}_\Gamma|^2 + K_2 |\nabla \times \mathbf{\Delta}_\Gamma|^2 + K_3 \left( |\partial_x \Delta_{\Gamma,1} - \partial_y \Delta_{\Gamma,2}|^2 + |\partial_x \Delta_{\Gamma,2} + \partial_y \Delta_{\Gamma,1}|^2 \right) \quad (12)$$

where $\Gamma = E', E''$. These terms can be understood as corresponding to the scalar, (axial) vector, and tensor components of the total derivative $\nabla \mathbf{\Delta}$, assuming no $z$ dependence. Fourier transforming this expression, it follows that, in the presence of an external symmetry-breaking field such as strain, the total gradient term is also twofold anisotropic[40].

**References:**

[S1]: C.-J. Lin, G.L. Gorman, Ch.H. Lee…H. Notarys, C.J. Chen. Magnetic and structural properties of Co/Pt multilayers. J. Magn. Magn. Mater **93**, 194 (1991).

[S2]: Borisenko, S. V. *et al.* Two energy gaps and fermi-surface 'arcs' in NbSe2. *Phys. Rev. Lett.* **102**, 166402 (2009).

[S3]: Sigrist, M. & Ueda, K. Phenomenological theory of unconventional superconductivity. *Rev. Mod. Phys.* **63**, 239–311 (1991).